\documentclass[12pt,preprint]{aastex}
\usepackage{psfig}

\shorttitle{Properties of mm galaxies}
\shortauthors{Dannerbauer et al.}

\begin{document}

\title{Properties of mm galaxies: Constraints from K-band blank fields \altaffilmark{1}}

\altaffiltext{1}{Based on observations collected at ESO, La Silla, Chile,
at the VLA, and on observations carried out with the IRAM Plateau de Bure
Interferometer.  IRAM is supported by INSU/CNRS (France), MPG (Germany)
and IGN (Spain).  The Very Large Array is a facility of the National Radio
Astronomy Observatory, which is operated by Associated Universities Inc.,
under cooperative agreement with the National Science Foundation.}

\author{H.~Dannerbauer\altaffilmark{2}, M.~D.~Lehnert\altaffilmark{2},
D.~Lutz\altaffilmark{2}, L.~Tacconi\altaffilmark{2},
F.~Bertoldi\altaffilmark{3}, C.~Carilli\altaffilmark{4},
R.~Genzel\altaffilmark{2} \& K.~Menten\altaffilmark{3}}

\altaffiltext{2}{Max-Planck-Institut f\"ur extraterrestrische Physik,
Postfach 1312, D-85741 Garching, Germany}

\altaffiltext{3}{Max-Planck-Institut f\"ur Radioastronomie, Auf dem
H\"ugel 69, 53121 Bonn, Germany}

\altaffiltext{4}{National Radio Astronomy Observatory, P.O. Box O,
Socorro, NM 87801, U.S.A.}



\begin{abstract}

We have used the IRAM Plateau de Bure mm interferometer to locate
with subarcsecond accuracy the dust emission of three of the brightest
1.2mm sources in the NTT Deep Field (NDF) selected from our 1.2mm MAMBO
survey at the IRAM 30m telescope. We combine these results with deep B
to K imaging and VLA interferometry.  Reliable identifications are an
essential step towards an understanding of the high redshift (sub)mm
galaxy population, towards testing the common belief that they are
scaled up analogs of local dusty ultraluminous galaxies, and in shedding
light on the possible connection to spheroid formation.  Strikingly,
none of the three accurately located mm galaxies MMJ120546-0741.5,
MMJ120539-0745.4, and MMJ120517-0743.1 has a K-band counterpart down to
the faint limit of K$_s$$>$21.9.  This implies that these three galaxies
are either extremely obscured and/or are at very high redshifts (z$\ga$4).
We combine our results with literature data for 11 more (sub)mm galaxies
that are identified with similar reliability. In terms of their K-band
properties, the sample divides into three roughly equal groups: (i)
undetected to K$\sim$22, (ii) detected in the near-infrared but not
the optical and (iii) detected in the optical with the possibility of
optical follow-up spectroscopy. We find a trend in this sample between
near-infrared to submm and submm to radio spectral indices, which in
comparison to spectral energy distributions (SEDs) of low redshift
infrared luminous galaxies suggests that the most plausible primary factor
causing the extreme near-infrared faintness of our objects is their high
redshift. We show that the near-infrared to radio SEDs of the sample are
inconsistent with SEDs that resemble local far-infrared cool galaxies
with moderate luminosities, which were proposed in some models of the
submm sky. We briefly discuss the implications of the results for our
understanding of galaxy formation.

\end{abstract}


\section{Introduction}\label{Introduction}

A local census of the distribution of the baryonic mass reveals, albeit
with large uncertainties, that of the baryons presently locked in stars,
a majority reside in spheroids \citep{per92,fuk98}. A key question then
is when and how all these baryons have come to reside in spheroids.
For several decades there have been two competing explanations.
The classical pictures are the 'monolithic collapse' of \citet{egg62}
versus the (hierarchical) merging model of \citet{sea78}.  With the
development of new techniques \citep[e.g., galaxies which 'drop-out'
in deep images;][]{ste96} and new technology (e.g., 10m class telescopes
and imaging mm/submm bolometers) our understanding of the cosmic history
of star-formation has accumulated rapidly and we are now in a position to
begin to address the question as to how and when spheroids were assembled.

A new route for the investigation of star-formation at the highest
redshifts (where spheroids probably have been assembled) has been opened 
by surveys of the dust reradiation in the submm/mm. Recently such surveys
have been carried out with the (sub)mm array cameras SCUBA and MAMBO
\citep[e.g., ][]{sma97,hug98,bar98,eales99,ber00a}. These investigations 
detect a 
population of luminous high redshift infrared galaxies that represent a 
significant fraction of
the cosmic submm background detected by COBE. A unique advantage is the
near constancy of observed (sub)mm flux with redshift because of the favorable
negative k-correction for galaxy SEDs peaking in the rest frame far-infrared.
As such, (sub)mm observations favor the detection
of objects resembling extreme versions of the most luminous infrared
sources in the local universe, ultraluminous infrared galaxies (ULIRGs) --
either powerful heavily obscured star-formation ($\sim$ 10$^2$ -- 10$^3$
M$_{\sun}$ yr$^{-1}$) or AGN.  Given their likely extreme luminosities
and the fact that a majority of the submm sources do not appear to be AGN
\citep[e.g.,][]{fab00,bau00,hor00,bar01a,bar01b,alm01}, it is hard to resist 
the speculation that a substantial fraction of faint (sub)mm population are 
massive spheroids in formation \citep[e.g.,][]{san99,fra99,lil99,tan99}.  
Analysis of the stellar content
of spheroids \citep[e.g.,][]{tra00a,tra00b} and investigations of the
dynamics of merging galaxies hosting powerful starbursts at low redshift
\citep{gen01} suggest diverse formation mechanisms and evolutionary
histories of early type galaxies.  If the speculation of the connection
between spheroid formation and the faint submm population turns out to be
well-founded, then determining the redshifts, bolometric luminosities,
and morphological and spectral properties of a significant number of
sources would greatly enhance our understanding of how spheroids formed
and which mechanism (monolithic collapse or merging) most strongly
influenced how and when spheroids formed. 

Beginning in the winter of 1998, our groups at the MPIfR, MPE, and
NRAO have been conducting a deep (rms $\sim$ 0.5mJy), wide (each of
three fields more than 100 arcmin$^{2}$) area survey at 1.2mm with
the Max-Planck-Millimeter Bolometer Array \citep[``MAMBO'';][]{kre98}
on the IRAM 30m telescope.  These surveys are specifically designed
to detect significant numbers of the brightest mJy-sources at 1.2mm
\citep{car99a,car00b,ber00a,ber00b}.  The large areal coverage and high
mapping speed of MAMBO have produced more than fifty firm detections of
bright mm sources (Bertoldi et al. 2002, in preparation), which directly
implies high luminosity through the strong negative k-correction for
sources that have spectral energy distributions peaking in the infrared.

One region that we have surveyed with MAMBO is a southern field centered
on, but larger than, the NTT Deep Field \citep[NDF;][]{arn99}.  Here we
report on mm interferometry, radio interferometry, and deep K-band and
BVRI imaging of three of the brightest likely nonlensed MAMBO mm sources
in the NDF (S$_{1.2mm}>$3mJy; for the perhaps more familiar S$_{850\mu
m}$ scale this converts to $\ga$8mJy for z$\sim$3 but depends on SED
and redshift). Our strategy is to identify possible counterparts and
to begin to assemble spectral energy distributions constraining the
nature and redshift of these objects. The interferometric observations
with the IRAM Plateau de Bure Interferometer (PdBI) and the VLA are
required to determine positions of subarcsecond accuracy and fluxes.
Several faint potential optical/near-IR counterparts are often present
inside or near the MAMBO beam for a mm galaxy \citep[e.g.,][]{lil99}. An
accurate location by interferometry is essential for any identification
because of this ambiguity and the possibility of `blank fields'. Without
that step, even the existence of potential counterparts does not imply
that the mm source is in fact associated with one of them.  We discuss our
results in the context of the data for other reliably identified sub(mm)
galaxies.  We adopt the cosmological parameters: $\Omega_{matter}=0.3$,
$\Omega_{\Lambda}=0.7$, and H$_0$=70 km s$^{-1}$ Mpc$^{-1}$.

\section{Observations, Data Reduction and Results}\label{ObsandRed}

\subsection{PdBI mm interferometry}\label{mminterobs}

We have observed three of the brightest (S$_{1.2 mm} > 3$ mJy) MAMBO
sources in the NTT Deep Field with the Plateau de Bure mm interferometer
\citep[PdBI;][]{gui92} in winter 2000/2001 for a determination of their
absolute positions, test for possible multiplicity or spatial extent,
and confirmation of the MAMBO fluxes.  The observations were performed in
the 5D configuration, with the receivers tuned to 238.46 GHz.  The phase
center was at the nominal MAMBO bolometer position and we integrated
between 12 and 22 hours per source, reaching an rms of between 0.9-1.2
mJy. We used 3C273 and 1124-186 to calibrate the temporal variation of
amplitude and phase, and 3C273 to calibrate the IF bandpass.  The absolute
flux scale was determined with observations of MWC349, CRL618 and 3C273,
and has an uncertainty of $\sim$20\%.  We calibrated and reduced the
data at IRAM, Grenoble using the GAG software packages CLIC and GRAPHIC.
The final naturally weighted dirty maps are shown in Fig.~\ref{pdbimaps}
along with the dirty beams. The FWHM of the beam for MMJ120546-0741.5
is 3.4\arcsec $\times$2.8\arcsec\ with PA=$-$21\arcdeg\ at 1.26 mm,
beams for the other sources are similar.

Two sources are detected at the $>$5$\sigma$ significance level
while MMJ120517-0743.1 is only tentatively detected (4.3$\sigma$) with
the PdBI. Fainter structure is seen closer to the nominal MAMBO
position of MMJ120517-0743.1 in both the PdBI and VLA maps but is
$<$3$\sigma$ in both. We
derived the positions and total 1.26mm flux densities listed in
Table~\ref{tab:pos} and Table~\ref{tab:flux} from point  source
fits to the calibrated visibilities in the UV plane. The positional
accuracy quoted in Table~\ref{tab:pos} is the quadrature sum of  the
statistical error of the point source fit with an astrometric systematic
uncertainty of 0.2\arcsec\ \citep[e.g.,][]{dow99}. The signal-to-noise
ratio of our data does not enable us to put strong constraints on
the source size at 1.26mm.  No significant indications for extension
are present in our $\sim 3$\arcsec\ beam data, however. The PdBI
map of MMJ120546-0741.5 suggests a possible second source near
12$^{h}$05$^{m}$46\fs7 -7\degr41\arcmin28\arcsec\ but its significance
is only  $\sim$3--4$\sigma$ and no additional supporting evidence is
given by a radio or optical/NIR counterpart.  Observations at 95GHz
(3.16mm) were obtained in parallel. None of  the sources is detected
at this frequency, consistent with dust emission.

\subsection{VLA 1.4GHz Interferometry}\label{vla}

The NTT deep field was observed at 1.4 GHz with the Very Large Array (VLA)
in April and May, 2001, for 15 hours in the B configuration (10 km maximum
baseline). Standard wide field imaging mode was employed (50 MHz total
bandwidth with two polarizations and 16 spectral channels).  The source
3C 286 was used for absolute gain calibration and 1224+035 was used for
phase and bandpass calibration.  The data were also self-calibrated
using sources in the NTT deep field itself.  Images were synthesized
and deconvolved using the wide field imaging capabilities in the AIPS
task IMAGR, and the primary beam correction was applied to the final
image. The full width at half maximum (FWHM) of the Gaussian CLEAN beam
was $7\arcsec \times 5\arcsec$ with a PA $= 0\arcdeg$. The rms noise
on the final image is between 13 and 15 $\mu$Jy over the 15\arcmin\
$\times$15\arcmin region covered by MAMBO.

Two of three MAMBO sources (MMJ120546-0741.5 and MMJ120539-0745.4)
have weak radio counterparts at the 3.4$\sigma$ and 4.3$\sigma$ level,
an upper limit is derived for MMJ120517-0743.1 (Table~\ref{tab:flux}).
The positional accuracy of these low S/N sources is dominated by the
statistical uncertainty of the fit, in Table~\ref{tab:pos} we adopt for
the positional uncertainty the commonly used measure $FWHM\over{2 S/N}$
\citep{Fomalont99}.

\subsection{Optical- and K-band Imaging}\label{Imaging}

Near-infrared/optical images of the NTT Deep Field publicly available at ESO
cover only the inner 20 to 30 arcmin$^2$ of the MAMBO field, e.g., a field 
of 16\,arcmin$^2$ to K$_{s}$=21.9 (3$\sigma$) for a 2\arcsec\ diameter aperture.
We imaged larger fields including the full area covered by MAMBO in 
B, V, R, and I at the ESO/MPI 2.2m telescope with the Wide-Field
Imager (WFI; \citet{baa98,baa99}),
and in K$_{s}$ at the ESO
NTT with the near-infrared imager/spectrometer SOFI \citep{moo98}.
The WFI data were taken during the spring of 2000 and 2001 while the
K-band images were taken in March 2001.  The field
size of the WFI is $34\arcmin\times33\arcmin$ while the field size for
individual SOFI frames is $4.9\arcmin\times4.9\arcmin$. We placed 8 contiguous
SOFI fields around the public K$_{s}$ image to cover a total field
of 13\farcm\/3$\times$13\farcm\/0. Both the WFI and SOFI frames
were dithered between exposures and the WFI was off-center (about 5')
from the NDF proper to avoid a nearby bright star.
During the WFI observations the conditions were mostly nonphotometric, 
while for the SOFI observations, all three nights were photometric.

The near-infrared data were reduced using IRAF and the ECLIPSE
package developed for reduction of ISAAC/SOFI data \citep{dev97}
and follow the standard reduction steps for near-infrared imaging data.
A preliminary field distortion correction was done using observations of the
cluster M10.
The WFI pipeline used for our reduction (Erben 2002, in preparation) 
includes the basic optical image reduction
steps (e.g.,  bias subtraction, flat-fielding, etc.) but also independent
astrometric solutions for each array in the WFI and allows the user to
interactively exclude images taken under cloudy conditions and/or bad
seeing through relative photometric and profile measurements of several
hundred stars per frame. 
The photometric zero-points for each filter were obtained through 
observations of the NICMOS standard stars \citep{persson98} for 
the K$_{s}$ images taken with SOFI, and through observations of
broad band standards \citep{landolt92} for the WFI data taken during 
photometric conditions. WFI data taken in nonphotometric conditions are tied
to that scale.
The 3$\sigma$ detection limits of our final images are B$=$27.4, V$=$26.6, 
R$=$26.2, I$=$25.3 and K$_{s}$=21.9 in 2\arcsec\ diameter apertures. 
The K-band images were aligned, residual distortions corrected, and rescaled 
to match the final optical images, using
the relative positions of about 100 sources per final K-band combined image
(of which there were 9 surrounding and including the public NDF K-band
image).  

\subsection{Astrometry}\label{Astrometry}
Estimating the astrometric uncertainty is crucial given the high density of 
faint potential counterparts to mm sources. Both the PdBI and VLA positions 
are obtained in the radio reference frame. The positions of the PdBI and
VLA detections for MMJ120546-0741.5 and MMJ120539-0745.4 (Table~\ref{tab:pos})
are consistent within
their errors which are 0.4\arcsec --0.5\arcsec\ for PdBI, and $\sim$1\arcsec\   
for the large beam VLA data of these faint sources.
For the optical/near-infrared data, the astrometry was accomplished
using several hundred stars from the USNO-A1 catalog spread across the 
34\arcmin$\times$33\arcmin\ field of view of the WFI. The rms deviation
of USNO-A1 input coordinates from the adopted WFI coordinate solution 
is 0\farcs45.

Systematic offsets are more difficult to establish, but several
comparisons can be made. Using common objects, the K$_s$ images were
aligned and  resampled to match closely the optical WFI images, any
systematic offsets are much less than one pixel of $\sim$0.2\arcsec . We
also compared the K-band (tied to USNO-A1 through the WFI image) and 
VLA frames using  33 well-detected VLA sources that had K-band sources
down  to K$_s$=21.9 within the likely positional uncertainty of each
 VLA source \citep[adopting as uncertainty the larger of 1\arcsec and
 3 $\times$ FWHM/(2$\times$S/N);][]{Fomalont99}.  The mean difference
between the near-infrared and VLA position is $0\farcs19\pm0\farcs07$ in
RA and $0\farcs23\pm0\farcs05$ in DEC. Optical/near-infrared positions
used later in this paper have been corrected for this small offset,
i.e. they should be on the radio system to a good approximation, with
random errors dominating systematic ones. 

From this analysis, we conclude that 1) differences
between the VLA and PdBI positions are consistent with the relatively low 
S/N of the VLA detections, and 2) the uncertainty of measured 
offsets between PdBI positions and near-infrared/optical objects, due to 
both systematic offsets and random errors,
is dominated by the random errors
of the individual measurements and typically $<$1\arcsec .

\section{All three mm sources are Blank Fields in the Optical/Near-Infrared}

Figure~\ref{fig:optnir} presents the near-infrared and optical images
for the regions around the three sources, with mm and radio positional
information overplotted. For MMJ120546-0741.5, no optical or near-infrared
sources are detected within the 11\arcsec\ MAMBO beam, which we adopt as
a conservative error circle estimate for the MAMBO bolometer map. The
closest near-infrared source is a galaxy to the south just outside
the MAMBO beam with K$_{s}$=20.3.  The mm and radio interferometric
observations locate MMJ120546-0741.5 at a position near the center of
the MAMBO beam, which is a `blank field' to the stringent limits given
in Table~\ref{tab:flux}, in particular to K$_{s}>$21.9 (3$\sigma$).
The nearest optical/near-IR sources are offset more than 5\arcsec\
from the interferometric position.

For MMJ120539-0745.4, there are three significant optical/NIR sources
within the MAMBO beam, and an identification lacking interferometric
observations probably would have picked one of them as the most likely
counterpart.  Strikingly, the interferometric observations demonstrate
MMJ120539-0745.4 to be a `blank field' as well (to limits listed in
Table~\ref{tab:flux}), with separations from the nearest optical/NIR
sources that are well above our estimated astrometric uncertainty.
The three optical/NIR sources are offset with respect to the PdBI
position by 2.4\arcsec\ to the south-west (K$_{s}$=20.5), 2.5\arcsec\
to the south-east (K$_{s}$=20.9), and 2.9\arcsec\ to the north-east
(K$_{s}$=20.4).

For the tentative detection of MMJ120517-0743.1, there is one significant
optical/NIR object near the eastern edge of the MAMBO beam, another
(not significant) object may be suggested closer to its center. The
PdBI source is just outside the nominal MAMBO beam but far off these two
objects and a `blank field' to e.g. K$_s$$>$21.9.  The optical/NIR object
closest to the PdBI position is a faint object (e.g. R=25.0) 3.4\arcsec\
further west outside the MAMBO beam.

We conclude that all three MAMBO sources are `blank fields' not having
K-band (or optical) counterparts down to very faint magnitudes. There
are indications that some mm galaxies may be very large and complex
objects \citep{ivi01,lut01}, however. This requires us to state this
conclusion precisely: Our astrometry shows that the region emitting
the mm (dust) emission does not have an optical/NIR counterpart down
to very faint limits. While the objects separated by $\ga$2.4\arcsec\
from MMJ120539-0745.4 are most likely unrelated foreground objects,
we cannot firmly exclude in the absence of redshift information a more
pronounced version of a situation similar to that of SMMJ14011+0252
\citep{ivi01}. In that source the optical emission regions, the radio
source that is obscured at optical wavelengths, and CO emission are
part of a large structure that extends in CO over $\ga$20kpc (taking
lensing magnification into account). The smaller radio and optical
peaks are offset relative to each other and to the CO peak by about a
third of that extent. In the case of MMJ120539-0745.4, assuming z$\sim$3
and our adopted cosmology, the angular separation of the nearby optical
objects from the mm position corresponds to $\ga$18kpc, i.e., the offset
would have to be significantly larger than that in SMMJ14011+0252. An
extended structure is, of course, hard to distinguish from yet another
configuration where the mm galaxy is part of an interacting system or
group with other, possibly optically more visible, galaxies present
nearby at the same redshift.

Unlike some other fields targeted by deep surveys, the NDF is not centered
on a massive moderate redshift cluster that provides amplification
by gravitational lensing. The wider field covered by our WFI imaging
includes the cluster [VFM98] 114 \citep[center 12$^{h}$06$^{m}$33\fs5
$-$07\degr44\arcmin28\arcsec]{vik98} at a redshift of 0.12.  None of
the three mm galaxies studied in this paper is close to this region,
however. The smaller scale morphology (Fig.~\ref{fig:optnir}) and the
faintness of nearby objects is not suggestive of lensing by individual
foreground objects or small groups. We hence assume in the following
that none of our objects are significantly amplified.

\section{Discussion}\label{disc}

\subsection{Are these z$\sim$3 Dusty Galaxies?}
\label{sect:arethesez3}

Starting with the first detections, SED-based arguments suggested
that submm galaxies are dusty luminous objects at z$\sim$2--4,
similar to scaled up versions of local ultraluminous infrared galaxies
\citep[e.g.,][]{hug98}. Much of the later evidence  for individual objects
\citep[e.g.,][]{lut01} and the statistical analysis on the basis of the
submm/radio redshift indicator \citep{car00a} both are consistent with
that notion. Progress has been limited, however, by (1) the small number
of reliable individual identifications, (2) the flattening and scatter
of the submm/radio redshift indicator which makes estimates at redshifts
above 2 fairly uncertain, and (3) the lack of observed far-infrared to
mm SEDs, which can help breaking the high redshift degeneracy of the
submm/radio redshift estimator by adding information on the location of
the rest frame far-infrared dust peak \citep[e.g.,][]{yun02}.

The dustiness of local ULIRGs creates red UV/optical SEDs with
considerable variation \citep{tre99}. Recently, significantly improved
rest frame UV to far-infrared SEDs have become available for a number
of local ULIRGs \citep{gol02,kla01}, which we use to analyze the
observed optical to mm SEDs of (sub)mm galaxies to place constraints
on their nature and redshift.  We have combined UV \citep{tre99,gol02}
and far-infrared \citep{kla01} photometry of ULIRGs with optical and
near-infrared photometry from the literature that we have approximately
corrected to similar (large) apertures, to obtain good observed UV to mm
SEDs for six ULIRGs: Arp 220, IRAS19254-7245, Mrk273, IRAS15250+0309,
IRAS12112+0305, and IRAS22491-1808. Figure~\ref{fig:ktomm} shows the
K-band magnitude predicted for a 5mJy source at 1.2mm, similar to our
targets, assuming these intrinsic ULIRG SEDs and redshifts up to 10.

If the three bright MAMBO sources have spectral energy distributions
similar to low redshift ULIRGs, then the faint K-band limits imply
that they are at very high redshifts $\ga$4 (Fig.~\ref{fig:ktomm}).
Alternatively, they may be at lower redshifts but more  obscured in the
intrinsic UV-optical than even the most obscured local ULIRGs  such as Arp
220. Both options are consistent with the radio to mm  redshift estimator
\citep{car99b,car00a}, given the VLA detections  (MMJ120546-0741.5 and
MMJ120539-0745.4) or limit (MMJ120517-0743.1). Considering measurement
errors, the scatter of the radio to mm spectral index vs. redshift
relation, and its flatness at high z, the three objects are consistent
with the radio/mm index at any redshift $\ga$2.  However, given their
near-infrared faintness, the three SEDs of the MAMBO sources are not
just scaled up SEDs of ULIRGs at z$\sim$3. We interpret this result in
the following in the context of the entire (sub)mm population.

\subsection{UV- to Radio SEDs and the Nature of Bright (sub)mm Sources}

We compare the properties of the MAMBO sources to a small sample of
(sub)mm galaxies that have been studied with similar methods. Some of
them do have optical counterparts (e.g., SMMJ14011+0252), indicating a
wider spread of properties. The sample used to improve the statistical
basis of our analysis includes only reliably located (sub)mm galaxies
from published blank field surveys and gravitational lens assisted
surveys. A key aspect for such a comparison are the criteria used for
identification. We consider as reliably located those objects with
available optical and near-infrared follow-up which are either (i)
detected in mm interferometry that directly probes the (sub)mm dust
emission (10 objects including the three sources presented here), or
(ii) detected at $>5\sigma$ in the (sub)mm and located through radio
interferometry. Radio identification is possible through the tight
radio-FIR correlation for star forming galaxies but can have problems,
as in the case of the identification of HDF850.1 \citep{dow99}. We
restrict radio identification here to sources well detected in the
(sub)mm, in order to limit contamination by galaxies with (sub)mm fluxes
that are highly uncertain because of Malmquist bias in the (sub)mm
surveys. We do not consider objects from radio-preselected surveys
\citep{bar00,cha01a,cha01b}. While such surveys efficiently locate a major
part of the (sub)mm galaxy population, they will be biased in redshift
distribution and thus the optical/near-IR properties of counterparts.
Table~\ref{tab:litsample} summarizes the properties of the 14 objects
meeting our selection criteria. By nature, this `reliable' sample from the
literature emphasizes the bright and luminous end of the (sub)mm galaxy
population, but is not a complete sample with strict selection criteria.

Optical/near-infrared counterparts of the (sub)mm galaxies are faint and
red. Only 4 of 14 are detected in the I band at a level where optical
spectroscopic follow-up is possible (I$\la$24). Two spectroscopic AGN
(with bright UV/optical  emission lines) are part of this group. In
addition, 4 to 5 of the 14 objects are detected in the near-infrared at
intermediate  levels of K$\sim$20--21 where detection is practical but
(near-infrared)  follow-up spectroscopy is very difficult. A significant
fraction of (sub)mm sources (5 to 6 of 14) is beyond the reach of
even deep K-band imaging to K$\sim$22. Each of the three groups 
represents about one third of the reliably identified (sub)mm galaxy
population. \citet{sma02} have classified submm galaxies from their
cluster  lensed sample into Class 0 (no counterpart to K$\sim$21,
I$\sim$26), Class I (detected at this depth in K but not I) and Class
II (detected in I), as  defined by  \citet{ivi00a}. They find that
$\sim$60\% of submm galaxies are Class 0. The small difference between
our findings and those of \citet{sma02} could  be due to the small
(though partly overlapping) samples, the less strict K-band limit
for their Class 0 (K=21)  than adopted here (K=22), and some of the 
\citet{sma02} Class 0 objects being of lower reliability in the original
SCUBA maps. It would be premature to interpret this modest variation
already in terms of our sample including only the most luminous objects,
or in terms of mm selection vs. submm selection.  Moreover, 
\citet{sma02}, find an indication from a sample of lensed SCUBA sources
that the faint (sub)mm galaxies (S$_{850\mu m}\la$4mJy) have a higher
fraction of K-band blank fields than the brighter (sub)mm sources. Our
observations, however, argue against such a trend.  Our sources, which are
likely to be non-lensed, are among the very brightest (sub)mm sources
if amplification is taken into account (Table~\ref{tab:litsample}).
This implies that a major fraction of (sub)mm sources, both "bright"
and "faint", are beyond the reach of optical and near-infrared
imaging and spectroscopy.

\citet{adel01} suggested that most of the galaxies forming the 850$\mu$m
cosmic background should be easily detectable in the rest-frame UV. As
discussed above, this is clearly not the case for the bright end of the
submm population (S$_{850\mu m}\ga$8mJy). Similarly reliable observations
and identification are currently impossible for the fainter (S$_{850\mu
m}\la$3mJy) objects comprising most of the 850$\mu$m background. If
those were easily detectable in the UV, however, the fainter and brighter
(sub)mm objects could no longer be linked as suggested by \citet{adel01},
i.e. through a strong correlation between dust opacity and star-formation
rate, but with the UV luminosity of all objects being fairly similar
irrespective of star-formation rate.

In the absence of spectroscopic redshifts for most objects, correlations
between observed properties are useful for constraining the redshift
and  evolution of submm sources, and their relation to local infrared
galaxies. Figure~\ref{fig:mmcorrs} shows diagnostic plots using
spectral indices and colors derived on the basis  of the optical
(I-band), near-infrared (K-band), submm, and radio properties  of this
sample. We compare these to the indices and colors expected for two
local  template SEDs moved to  redshifts between 1 and 5: Arp220 as
an example of a  highly obscured local ultraluminous infrared galaxy
and M82 as an example  of a less obscured luminous infrared galaxy
(LIRG). The non-detection of many objects in K and, in particular, I
makes the two right panels of Fig.~\ref{fig:mmcorrs} less constraining
on the nature of these sources. It is clear that some submm galaxies are
redder in I-K than M82 placed at any plausible  redshift, and one might
speculate that the objects undetected in both I and K  are similarly
red. At least a subset of (sub)mm galaxies is redder in the rest  frame
UV/optical/near-IR than a normal dusty starburst like M82, and is better
represented by an Arp\,220--like SED.

The top left panel of Fig.~\ref{fig:mmcorrs} combines two spectral
indices: On the abscissa the submm/radio spectral index frequently used
as a redshift  indicator, and a spectral index between observed K-band
and 850$\mu$m, i.e., rest frame optical and submm. The rest frame optical
emission of the observed (sub)mm sources is very  weak compared to the
submm/far-infrared one, clearly requiring a high dust content. Even a
local dusty starburst like M82 is not obscured enough to represent the
majority of (sub)mm objects. A highly obscured ULIRG like Arp220 provides
a better fit, but on average the obscuration of (sub)mm objects may be yet
larger (see also Sect.~\ref{sect:arethesez3}). A trend is seen  between
the two indices whereby the sources with the steepest 850$\mu$m to 1.4GHz
spectral indices also tend to be the sources with the steepest 2.2$\mu$m
to 850$\mu$m spectral indices. This indicates, as perhaps expected, that
the K-band to submm index significantly depends on redshift because
of the widely differing k-corrections at its defining frequencies,
and that variations in source properties do not fully overwhelm this
effect. Interestingly, the objects that are blank fields down to very
faint K magnitudes all cluster in the (lower) right of this diagram,
i.e. where the radio to submm redshift indicator puts them at the
higher end of the redshift spread of (sub)mm galaxies.  This result is
most simply explained by one of the two explanations for K-band blank
fields put forward in the previous section:  Most plausibly these are
the highest redshift objects.  The submm/radio redshift indicator
is unfortunately not accurate enough to assign good redshifts, but
distinguishes the K-band faint objects from the rest of the population.
If they were yet more extremely obscured objects than ULIRGs but with a
redshift range similar to the other sources in Fig.~\ref{fig:mmcorrs},
the scatter across the lower half of the diagram would likely be greater
than observed.  Moreover, having a significant fraction of (sub)mm sources
at high redshift apparently does not violate any constraint from the SED
of the  cosmic infrared background (CIRB).  For example, \citet{gis00}
showed that the relative contribution of objects with spectral energy
distributions like local ultra-luminous infrared galaxies to the CIRB
at high redshifts (z$>$4) can be quite significant ($\approx$20-40\%)
without violating any constraints provided by its spectrum.

All galaxies detected in (sub)mm surveys, except perhaps for a few of
the very highest redshift ones, will be observed on the long wavelength
side of their rest frame far-infrared emission maximum, while local
comparison objects were selected by IRAS near the maximum or slightly to
its short wavelength side. This situation can induce a bias, favoring
detection of objects with somewhat colder dust in the (sub)mm surveys.
Trying to match the observed number counts, cosmological simulations
by \citet{fardal01} invoke colder SEDs more similar to local luminous
infrared galaxies \citep{dun00} rather than the warmer SEDs of local
ULIRGs \citep[see also][]{eales00}. Keeping the observed (sub)mm flux
fixed, a colder SED will result in a lower redshift and/or less luminous
object. The additional optical/near-infrared and radio constraints can
test the plausibility of these assumptions.  In Fig.~\ref{fig:dunnecorrs},
we show where the colder (T$<$35K) among the local galaxies observed by
\citet{dun00} would be placed for several redshifts on a diagram combining
near-infrared/submm and submm/radio spectral indices, as already shown
for the (sub)mm sources in Fig.~\ref{fig:mmcorrs}. Clearly those colder
LIRG-like SEDs predict K-band magnitudes that are too bright and are not
consistent with the results for the (sub)mm galaxies. The scenario of
\citet{fardal01} where the SEDs of submm galaxies are closer to the cooler
LIRGs studied by Dunne et al.  rather than the ULIRGs adopted by other
workers is, hence, inconsistent with the overall SED information including
the observed near-infrared (rest frame optical). This is a consequence of
the lower luminosity and colder objects in the local universe having a
higher relative fraction of optical to infrared luminosity. Suppressing
the K-band emission observed at higher redshift for such SEDs to the
required level would require that the (sub)mm galaxies are at even higher
redshifts than discussed above.  If the rest frame far-infrared SEDs of
(sub)mm galaxies are cool, their overall SEDs still must be distinct
from local universe cold LIRGs in having a much stronger ULIRG-like
obscuration of optical emission. Such galaxies - without analogs in the
local universe - could be envisaged as a collection of the `coldest' and
most obscured star forming regions like SgrB2, without admixture of the
warmer and less obscured regions normally representing other evolutionary
stages of star forming complexes.  Alternatively, they could be modelled
as dilute obscured `cirrus'-type systems (Efstathiou \& Rowan-Robinson,
in preparation), but only with the stringent assumption of constant
optical depth dust enclosing the entire large region.

\subsection{Implications}

We have discussed two possibilities for the nature of bright (sub)mm 
sources that are very faint in the near-infrared (K$\ga$22). They are
either heavily obscured starbursts/AGN -- in this context more heavily
obscured than any well-studied local starbursts or QSOs -- or are, as 
suggested by the analysis given in the previous section, similar
to low redshift dusty powerful starbursts but located at very high redshifts
(z$\ga$4).  In either case, several important inferences can be drawn.

In the local ULIRG population, the amount of UV light observed,
even if corrected for extinction, traces only a small fraction of
the bolometric luminosity \citep[$\la$10\% ;][]{gol02}.  This implies
that most of the intrinsic UV emission is not detected in ULIRGs but is
completely obscured. The observed UV emission simply cannot be related to
the sites where most of the bolometric luminosity is produced.  To make
the rest-frame UV/optical to far-infrared SED even redder than the most
highly reddened galaxies in the local universe would require special
circumstances in order to lower the ratio of optically visible to total
luminosity yet further.  One such special circumstance is an extremely
young starburst which is completely obscured in all directions. To
maintain heavy obscuration the starburst must be younger than the time
necessary for significant gas removal due to the feedback from stellar
winds and supernovae.  This timescale is at most a few $\times$ 10$^7$
years and likely significantly less \citep{leh99,hec00,tho00}.  Otherwise,
in a second special case, the gas accretion rate and timescale must be
such that as the feedback clears regions of young or intermediate age
stars, there would be inflowing gas available to maintain the overall
obscuration.  Maintaining such extremely high obscuration would require
gas with substantial amounts of dust to fall onto the region of intense
star-formation and have a distribution such that there are very few
regions of low optical depth in the rest-frame UV. In addition, such
vigorous infall must be maintained for at least the gas clearance time
scale and likely much longer.  A third possibility would be to invoke
a large scale but still very thick and non-starforming cocoon around
the star forming zone, delaying destruction by gas removal but putting
strong requirements on size and uniformity of such a cocoon.

The other plausible scenario for the nature of these bright (sub)mm
sources is that their SEDs are similar to the most obscured local
dusty starbursts but are at extremely high redshift (z$\ga$4).  It is
now well-understood that local ULIRGs are a result of the interaction
or merger of two gas rich spiral galaxies \citep[see][and references
therein]{SandM96}.  Continuing this analogy of (sub)mm sources being
high redshift galaxies with extremely high star-formation rates and SEDs
that are ULIRG-like would then suggest that they too are the merger of
massive disk galaxies. Higher gas fractions in high redshift galaxies
may support such a process. The main implication of this scenario is
the need to have massive units in place for the z$>$4 merger, gas-rich
enough to sustain star formation rates of the order 1000\,M$_\sun$ yr$^{-1}$
and provide the material obscuring this burst.

In either event, the implications for our understanding of the formation
of spheroids is interesting.  Requiring extremely high obscuration to
explain the properties of the brightest mm sources is consistent with
a model whereby spheroids are formed through a process similar to the
``monolithic collapse'' model. That is, through the rapid collapse of
large amounts of gas which is maintained over several to many dynamical
time scales \citep[of-order 100 Myrs -- 1 Gyr; see e.g.,][]{jimenez99}.
A model like this has the attractive feature of being able to supply
the gas necessary to both maintain the high UV extinction and the large
fraction of the total bolometric luminosity that must be emitted in
the infrared/submm.  If, on the other hand, the SEDs of the mm sources
are similar to the local ULIRGs, then the problem is that at z=4 (the
low end of possible redshifts for these sources, see Fig.~3), the age
of the Universe is only about 1.5 Gyr or about 11\% of its current age.
Within the context of the popular hierarchical merging models, this time
is too short to form massive metal-rich disks for the necessary merging to
take place \citep[see][]{fardal01}.  Hierarchical merging models simply
have great difficulty in producing the high rates of star-formation
which these galaxies likely have at such high redshifts.  For example,
in the results of the simulation by \citet{fardal01}, the sources that
can be identified with the faint submm population (S$_{850\mu m}$ $\ge$
1 mJy) are found to redshifts of about 1--3 with no sources at z$\ga$4
obtaining a large enough luminosity sufficient to be detected at the
fluxes of the source investigated here.  In both models, merging and
``monolithic collapse'', prompt metal enrichment may also be necessary to
create sufficient amounts of dust to effectively allow for the virtually
complete re-emission of the bolometric luminosity into the infrared/submm.

\section{Conclusions}\label{Conclusions}

We have used mm and radio interferometry to locate accurately three of
the brightest and likely non-lensed MAMBO mm sources -- MMJ120546-0741.5,
MMJ120539-0745.4, and MMJ120517-0743.1.  These objects are among the
very brightest (sub)mm galaxies if amplification is taken into account.
Associated deep optical and K-band follow-up shows that these sources
do not have optical or near-infrared counterparts down to very faint
magnitudes (K$_{s}$$>$21.9).  The near-infrared faintness of these
bright mm galaxies implies that they must either be at very high redshift
(z$\ga$4) or more highly obscured than even the dustiest local ULIRGs.

We have combined these limits with results from the literature for those
among the brightest (sub)mm sources that we consider located reliably
to subarcsecond accuracy. About a third each of the total sample are
(1) detected in the optical with follow-up possibility, (2) optically
faint but detected at K$\sim$20-21, and (3) undetected in K at levels
similar to the three MAMBO galaxies (K$\ga$22). The combined optical
-- near-infrared -- submm -- radio data for the sample suggest the
near-infrared faintness of our objects to be most likely due to very
high redshifts and not due to obscuration alone.  The SEDs of local
galaxies that are cool in the far-infrared do not explain the overall
characteristics of these galaxies very well and therefore models of
this type are not favored on the basis of these data.  The nondetection
of a major fraction of the (sub)mm population in the near-infrared
and optical stresses the need for future new methods like wideband CO
searches, mid-infrared emission features, or maser lines in obtaining
accurate redshifts for this important group of high redshift objects,
in order to understand the (sub)mm galaxy population as a whole.

The existence of a sizeable fraction of K-band faint (sub)mm galaxies has
implications for our understanding of the formation of galaxies.
Specifically, such results will test whether the faint mm population is
similar to the merger-induced dusty powerful starbursts in the local
universe, or if they represent something more extreme -- e.g., the
``monolithic collapse'' of massive spheroids.  If the brightest mm sources
lie mostly at high redshift (z$\ga$4) and have extreme luminosities
(L$_{bol}$$>$ few $\times$ 10$^{12}$ L$_{\sun}$), then hierarchical
merging models will have difficulty in explaining these sources within
the general context of galaxy formation and evolution.  As such, these
bright source may provide a critical test of these widely favored models.

\acknowledgements{We would like to thank the staffs of ESO, IRAM, and VLA
for their support with the data acquisition and reduction. Additionally,
we are grateful to the IAEF Bonn and the ESO EIS Survey for providing
us with the necessary hardware for reducing the WFI data. Special thanks
go to M. Schirmer and T. Erben for help in the WFI reduction and making
available the WFI pipeline. Finally, we appreciate the comments of the
anonymous referee which improved this paper. This research has made
use of the NASA/IPAC Extragalactic Database (NED) which is operated
by the Jet Propulsion Laboratory, California Institute of Technology,
under contract with the National Aeronautics and Space Administration.}

%
%

%
%

\clearpage

\begin{deluxetable}{llll}
\tabletypesize{\small}
\tablecolumns{4}
\tablewidth{0pt}
\tablecaption{Positions of the three NDF mm galaxies}
\tablehead{
\colhead{Object}&\colhead{Instrument}&
\colhead{RA (J2000)}&\colhead{DEC (J2000)}
}
\startdata
MMJ120546-0741.5&MAMBO&12$^{h}$05$^{m}$46\fs57$\pm0\fs20$&$-$07\degr41\arcmin33\farcs4$\pm$3\arcsec\\ 
               &PdBI &12$^{h}$05$^{m}$46\fs59$\pm0\fs02$&$-$07\degr41\arcmin34\farcs3$\pm$0\farcs37 \\
               &VLA  &12$^{h}$05$^{m}$46\fs55$\pm0\fs05$&$-$07\degr41\arcmin33\farcs2$\pm$1\farcs03 \\
MMJ120539-0745.4&MAMBO&12$^{h}$05$^{m}$39\fs32$\pm0\fs20$&$-$07\degr45\arcmin26\farcs4$\pm$3\arcsec\\
               &PdBI &12$^{h}$05$^{m}$39\fs47$\pm0\fs02$&$-$07\degr45\arcmin27\farcs0$\pm$0\farcs33 \\
               &VLA  &12$^{h}$05$^{m}$39\fs45$\pm0\fs04$&$-$07\degr45\arcmin26\farcs1$\pm$0\farcs81 \\
MMJ120517-0743.1&MAMBO&12$^{h}$05$^{m}$17\fs93$\pm0\fs20$&$-$07\degr43\arcmin06\farcs6$\pm$3\arcsec \\
               &PdBI &12$^{h}$05$^{m}$17\fs59$\pm0\fs03$&$-$07\degr43\arcmin11\farcs5$\pm$0\farcs51\\
               &VLA  &...                        &...                \\
\enddata
\label{tab:pos}
\end{deluxetable}

%
%

\clearpage

\begin{deluxetable}{lccccl}
\tabletypesize{\small}
\tablecolumns{6}
\tablewidth{0pt}
\tablecaption{Fluxes of the three NDF mm galaxies}
\tablehead{
\colhead{Band}&\colhead{Unit}&
\colhead{MMJ120546-0741.5}&\colhead{MMJ120539-0745.4}&
\colhead{MMJ120517-0743.1}&\colhead{Comment}\\
\colhead{(1)}&\colhead{(2)}&
\colhead{(3)}&\colhead{(4)}&
\colhead{(5)}&\colhead{(6)}}
\startdata
m$_B$        &    mag&$>$27.4&$>$27.4&$>$27.4 &WFI\\
m$_V$        &    mag&$>$26.6&$>$26.6&$>$26.6 &WFI\\
m$_R$        &    mag&$>$26.2&$>$26.2&$>$26.2 &WFI\\
m$_I$        &    mag&$>$25.3&$>$25.3&$>$25.3 &WFI\\
m$_J$        &    mag&  ...  &...    &$>$23.9 &SOFI (p)\\
m$_{K_s}$    &    mag&$>$21.9&$>$21.9&$>$21.9 &SOFI (p)\\
S$_{1.2mm}$  &    mJy&6.5$\pm$0.8&3.5$\pm$0.8&4.4$\pm$0.8 &MAMBO\\
S$_{1.26mm}$ &    mJy&6.1$\pm$1.2&6.4$\pm$1.0&3.9$\pm$0.9 &PdBI\\
S$_{3.16mm}$ &    mJy&$<$0.84&$<$0.39&$<$0.60 &PdBI\\
S$_{1.4GHz}$ &$\mu$Jy&44$\pm$13&56$\pm$13&$<$39&VLA
\tablecomments{
Col. (1) --- Band in which flux is measured.  
Col (2) --- Units of the flux density measurements.
Col. (3)-(5) --- Measurements for the three objects. Optical/near-IR
magnitudes were measured in a 2\arcsec\ diameter aperture, and are on the `Vega' 
system. Limits are 3$\sigma$.
Col. (6) --- Comments. (p) = Limits for MMJ120517-0743.1 from public NTT deep field data, 
http://www.eso.org/science/sofi\_deep/index.html.}

\enddata
\label{tab:flux}
\end{deluxetable}

\clearpage

%
%

\begin{deluxetable}{llcccrrrrr}
\tabletypesize{\small}
\tablecolumns{10}
\tablewidth{0pt}
\tablecaption{Submm/mm sources with interferometric confirmation}
\tablehead{
\colhead{Source}     &\colhead{Alias}&\colhead{z}&
\colhead{Refer.}       &\colhead{Ident}&
\colhead{Lens}       &\colhead{S$_{850}$}&
\colhead{S$_{1.4}$}  &\colhead{K}&
\colhead{I}
\\
\colhead{}           &\colhead{}&\colhead{}&
\colhead{}           &\colhead{}&
\colhead{}           &\colhead{mJy}&        
\colhead{uJy}        &\colhead{mag}&
\colhead{mag}        
\\
\colhead{(1)}&\colhead{(2)}&
\colhead{(3)}&\colhead{(4)}&
\colhead{(5)}&\colhead{(6)}&
\colhead{(7)}&\colhead{(8)}&
\colhead{(9)}&\colhead{(10)}
}
\startdata
 00266+1708&          &    &f00        &mm,ra&   2.4&18.6&   94&22.45 &$>$26.1\\
 02399-0134&          &1.06&sou99,s00  &ra   &   2.5&11.0&  500&16.34 &19.31\\
 02399-0136&L1+L2     &2.81&i98,f98    &mm,ra&   2.5&25.4&  526&19.01 &21.73\\
 09429+4658&H5        &    &s99,s00    &ra   &   2.0&17.2&   32&19.58 &$>$25.6\\
 10520+5724&L850.1    &    &l01        &mm,ra&     1&10.5&   74&19.80 &$>$26.0\\
120517-0743.1&        &    &           &mm   &    1&(9.8)&$<$39&$>$21.9&$>$25.3\\  
120539-0745.4&        &    &           &mm,ra&     1&(16.0)& 56&$>$21.9&$>$25.3\\
120546-0741.5&        &    &           &mm,ra&     1&(15.0)& 44&$>$21.9&$>$25.3 \\
 12368+6212&HDF850.1  &    &d99,m00    &mm,ra&  $>$1& 7.0&   20&$>$22.0&$>$26.1\\
 14009+0252&J5        &    &i00        &ra   &   1.5&15.6&  529&20.96 &$>$23.0\\
 14011+0252&J1        &2.56&i00,i01    &mm,ra&   3.0&14.6&  115&17.97 &21.10\\
 14171+5229&CFRS14A   &    &b00a,g00   &mm,ra&     1& 8.8&  120&19.66 &24.13\\
154127+6615&A2125-2   &    &b00a,d03   &mm,ra&$\ga$1&(10.0)& 81&$>$21.2&$>$23.9\\
154127+6616&          &    &b00a       &ra   &$\ga$1&(10.5)& 67&20.50 &$>$23.5\\
\tablecomments{
Col. (1,2) --- Name and aliases.
Col. (3) --- Spectroscopic redshift where available.
Col. (4) --- References.
Col. (5) --- Identification method: mm and/or radio interferometry.
Col. (6) --- Lensing magnification. {\em Subsequent quantities
             are as observed, i.e. not corrected for lensing}.
Col. (7) --- 850$\mu$m flux density () means estimated as 
             2.5$\times$S$_{1.2mm}$.
Col. (8) --- 1.4GHz flux density.
Col. (9,10) --- Magnitudes of suggested near-infrared/optical counterpart 
               (Vega).
}
\tablecomments{
Remarks on individual objects: 12368+6212 (HDF850.1): originally identified 
by d99 with a faint optical arc. We follow m01 who reject this identification
on the basis of improved relative astrometry, and adopt parameters of the
arc as upper limits. 1.4GHz flux estimated from 8.5GHz flux. The object may 
be lensed. 14171+5229: 1.4GHz flux estimated from 5GHz flux. 154127+6615,
154127+6616: I band limits estimated from R band limits.
}
\tablecomments{
References:
b00a: Bertoldi et al. 2000a ---
d99: Downes et al. 1999 ---
d03: Dannerbauer 2003 ---
f98,f00: Frayer et al. 1998, 2000 ---
g00: Gear et al. 2000 ---
h98: Hughes et al. 1998 ---
i98,i00,i01: Ivison et al. 1998, 2000, 2001 ---
l01: Lutz et al. 2001 ---
m01: Muxlow et al. 2001 ---
s99,s00: Smail et al. 1999, 2000 ---
sou99: Soucail et al. 1999.
}
\enddata
\label{tab:litsample}
\end{deluxetable}

\clearpage

\centerline{\bf Figure Captions}
\figcaption[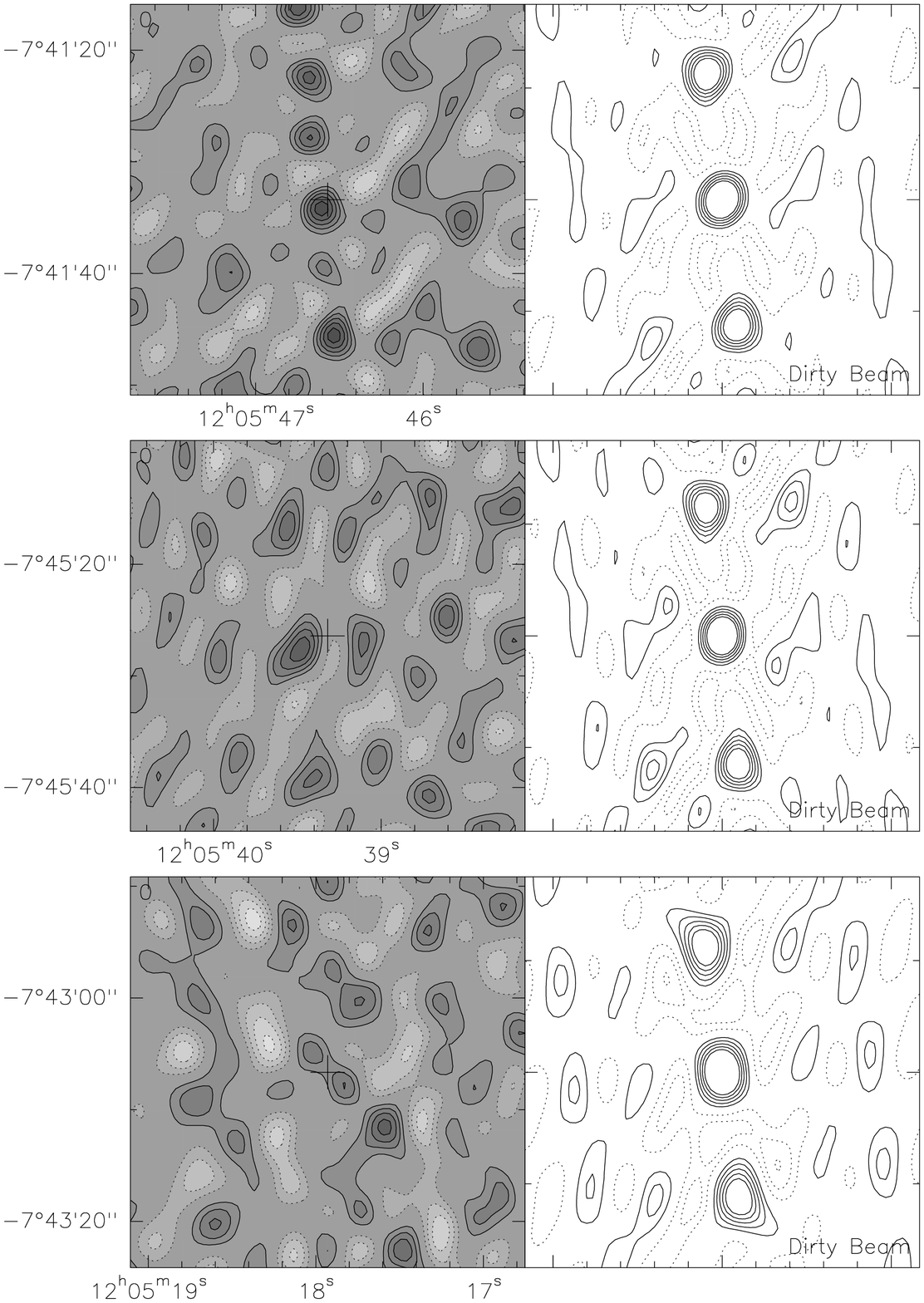]
{`Dirty' 1.26mm interferometric maps for the three PdBI detections.
From top to bottom: MMJ120546-0741.5, MMJ120539-0745.4, and
MMJ120517-0743.1. Each image is $35\arcsec \times 35\arcsec $ and oriented
such that north is at the top and east is to the left. The plus indicates
the phase center, placed at the position of the mm source originally
estimated from the bolometer map. The contour levels are spaced by
1 mJy and the solid and dotted lines represent positive and negative
contours respectively.  The dirty beams are shown to the right, they
show considerable side lobes because of the equatorial declination. The
beam profiles for the three sources are similar because of the similar
declination and UV coverage.
\label{pdbimaps}}
\figcaption[f2.ps]
{K, I, and R band images (left to right) for the fields of the three
MAMBO sources, from top to bottom MMJ120546-0741.5, MMJ120539-0745.4,
and MMJ120517-0743.1.  Size of each image and orientation is as in
Fig.~\ref{pdbimaps}.  The large circles represent the size of the MAMBO
beam ($\sim$11\arcsec\ diameter). Small crosses indicate the position and
2$\sigma$ errors of the PdBI detection, and the larger crosses positions
and 2$\sigma$ errors obtained from VLA 1.4 GHz interferometry. No
near-infrared or optical counterparts are detected in these deep images,
in particular to K$_s$=21.9.
\label{fig:optnir}}
\figcaption[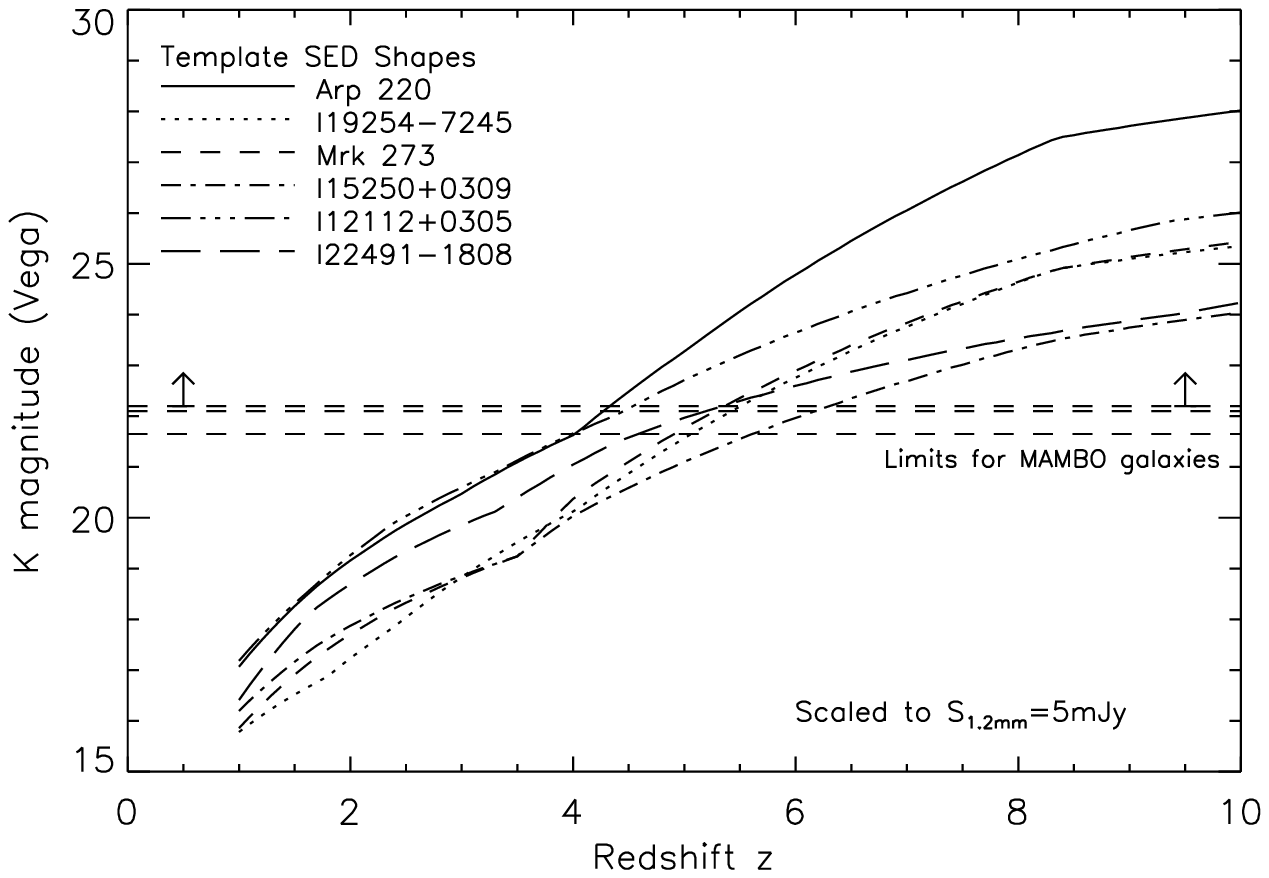]{Expected K-band magnitude (Vega) of a mm galaxy as a
function of redshift. The rest frame SED is assumed to have the shape of
local ULIRG SEDs for which high quality UV to FIR photometry is available.
The redshifted SEDs were scaled to S$_{1.2mm}$=5\,mJy, representative
for our MAMBO sources.  Line styles identify the local ULIRGS used as SED
templates.  Horizontal dashed lines indicate the 3$\sigma$ limits for the
three MAMBO galaxies, again scaled to match a common S$_{1.2mm}$=5\,mJy.
\label{fig:ktomm}}
\figcaption[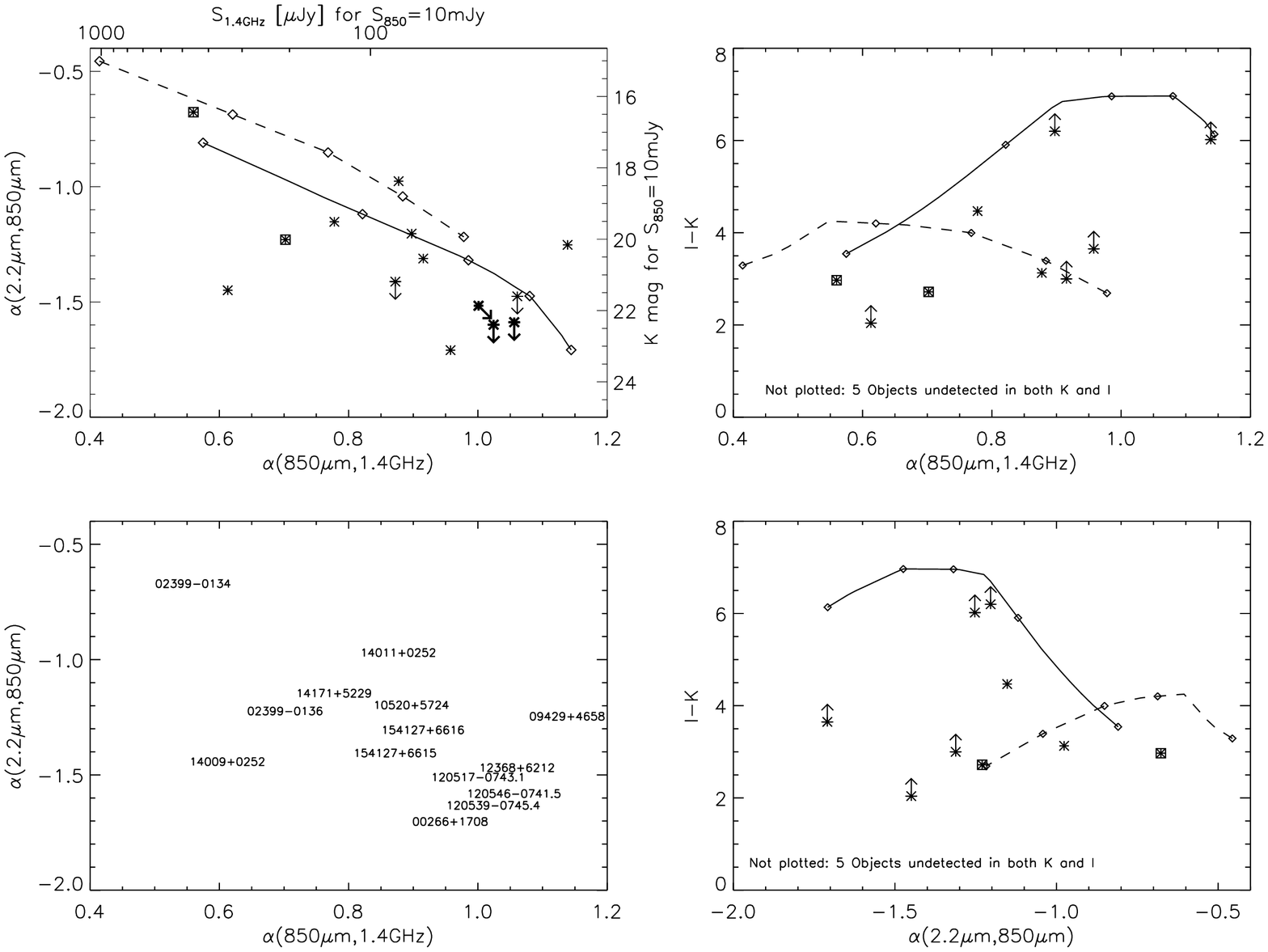]{Projections of a colour space for (sub)mm 
galaxies derived on the basis of radio, submm, K-band, and I-band data.
The top left panel shows the relation of the submm -- radio spectral index
and a K-band submm spectral index. 
The top and right axis of this panel illustrate
the corresponding radio and K-band brightness of a fiducial bright submm source.
Two spectroscopic AGN are marked by squares, and the three sources presented
in this paper are highlighted by thick symbols on the lower right. 
Overplotted lines show the loci of Arp220 (continuous) and M82 (dashed) for 
redshifts of 1 (left) to 5 (right).
The bottom left panel repeats the same information
with source names indicated. The top right and bottom right panels combine
the colour I-K with the submm/radio and K-band/submm spectral indices.
\label{fig:mmcorrs}}
\figcaption[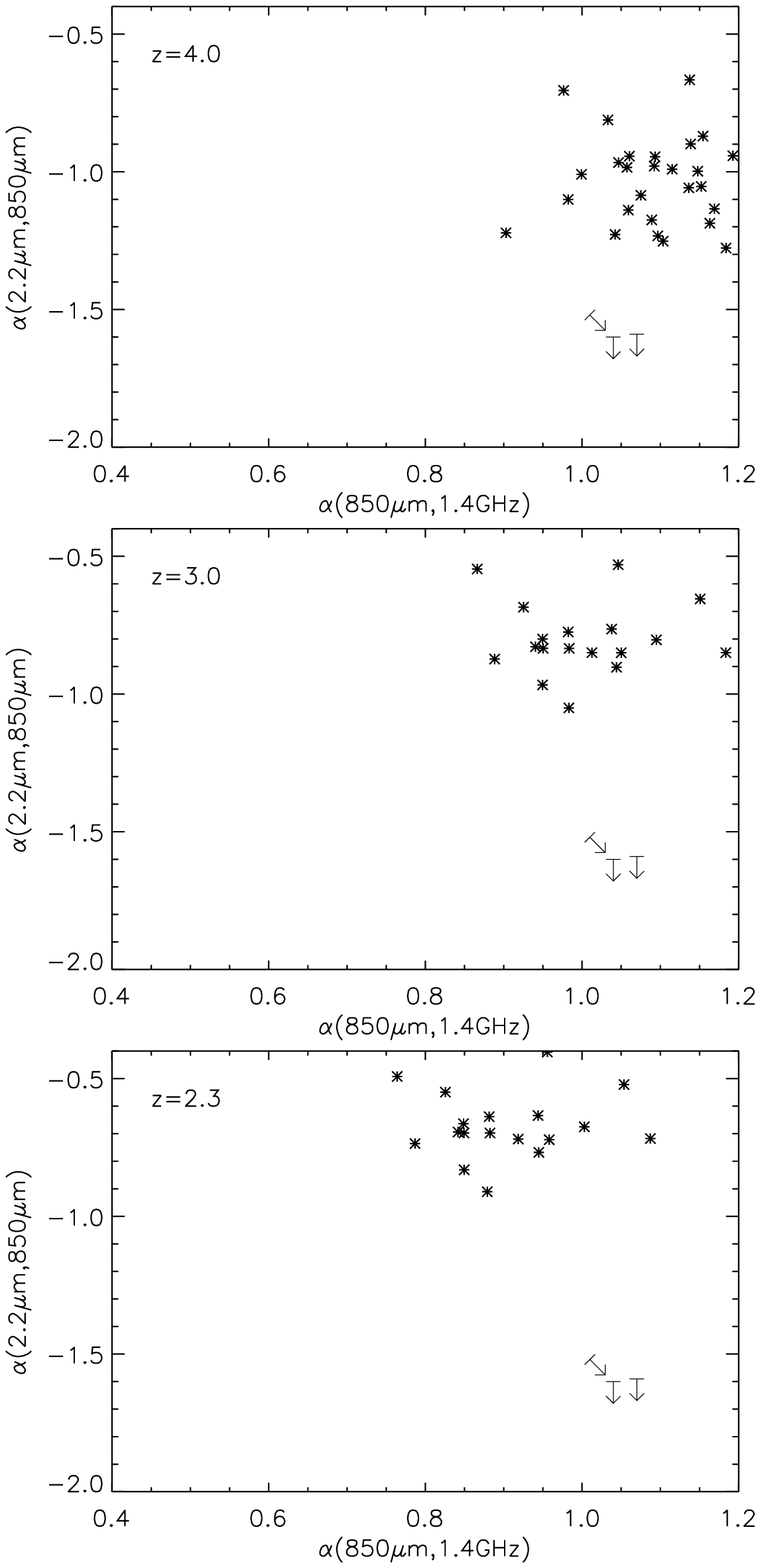]{Near-infrared/submm vs. submm/radio
spectral indices for cool (T$<$35K) galaxies from the \cite{dun00} sample of
local galaxies shifted to three different redshifts. The indices for these 
galaxies (asterisks) and z$\sim$4 (top), z$\sim$3 (middle), and 
z$\sim$2.3 (bottom) are derived on the basis of B,V,R fluxes
(which correspond to K-band at those redshifts) for those Dunne et al. 
galaxies where they are available (B,V) or can be extrapolated (R), and the 
submm and 1.4GHz
fluxes corrected for redshift using the Dunne et al. blackbody fits (submm)
and a spectral index of -0.8 (radio). For comparison, we add the limits for the
three MAMBO galaxies in each panel. For all these redshifts, the Dunne et al.
objects
are too bright in K to match the MAMBO galaxies and the other (sub)mm 
galaxies shown in the same diagram in the top left panel of 
Fig.\ref{fig:mmcorrs}. 
\label{fig:dunnecorrs}}
%

\newpage
\epsscale{0.8}
\plotone{f1.ps}
\newpage
\epsscale{1.0}
\plotone{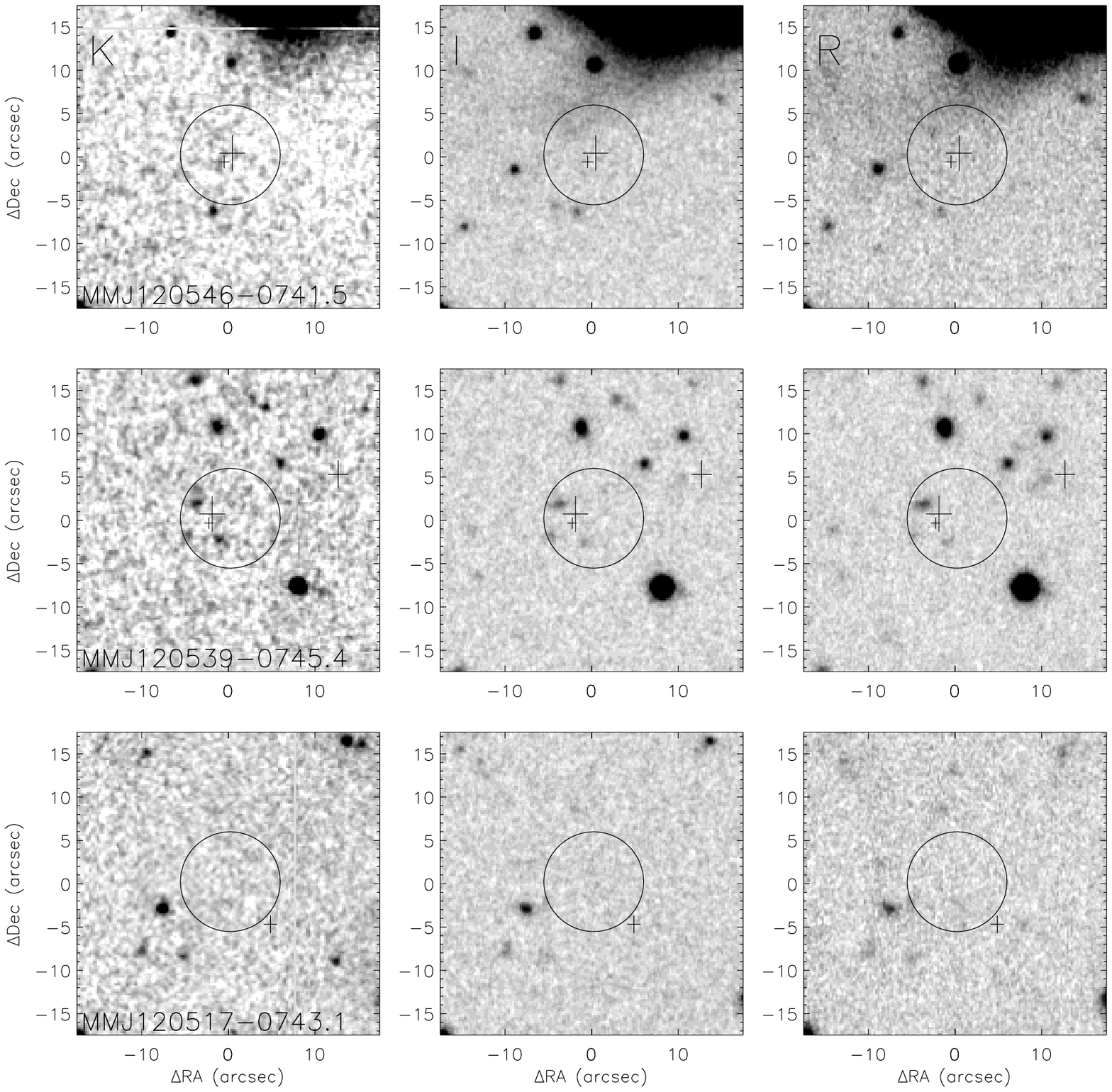}
\newpage
\plotone{f3.eps}
\newpage
\plotone{f4.eps}
\newpage
\epsscale{0.6}
\plotone{f5.eps}

\end{document}